\documentstyle[preprint,aps]{revtex}
\tighten
\begin{document}
\draft

\preprint{\hfil{NCL94--TP10}}% ADP-94-9/M23
\title{ Conformal anomalies on Einstein spaces with Boundary}
\author{Ian G. Moss}
\address{
Department of Physics, University of Newcastle Upon Tyne, NE1 7RU U.K.
}
\author{Stephen J. Poletti}
\address{
Department of Physics and Mathematical Physics,
 University of Adelaide, South Australia.
}
\date{May 1994}
\maketitle
\begin{abstract}
The anomalous rescaling for antisymmetric tensor fields, including
gauge bosons, and Dirac fermions on Einstein spaces with boundary has
been prone to errors and these are corrected here. The explicit
calculations lead to some interesting identities that indicate a deeper
underlying structure.
\end{abstract}

\section{INTRODUCTION}

In this letter we would like to correct some results that we gave
previously for the conformal anomaly of quantum fields on spaces with
boundaries \cite{moss}. These results caused some consternation because
there seemed to be a disagreement between general results obtained
from heat kernel asymptotics \cite{moss,moss2,poletti} and direct
calculations of the anomaly
\cite{death,death2,e,luoko,b,b2,k,k2,k3,e2}. We can confirm now that
the situation has been resolved with the discovery by Vassilevich
\cite{v} of corrections to the heat kernel asymptotics
\cite{mk,kennedy,moss3,moss4,branson}.

The physical motivation for these calculations is connected with
quantum cosmology and the quantum state of the universe, where boundary
effects play an important role. In the path integral approach boundary
terms are important in one--loop corrections.

Besides the practical applications of these results there are also a
number of mathematical coincidences that seem to 1)      originate in an
interesting identity. Before discussing these results we will use the
calculation of the conformal anomaly of a Dirac fermion as a
consistency check.

\section{ONE LOOP AMPLITUDES}

It is convenient to define quantum amplitudes using a path integral
over field configurations on a Riemannian manifold ${\cal M}$ with
boundary $\Sigma$,
\begin{equation}
e^{-\Gamma}=\int d\mu[\phi]\,e^{-S[\phi]},
\end{equation}
where $S[\phi]$ is the Riemannian action and $\hbar=1$.

In the semi--classical approximation the path integral is dominated by
saddle point configurations $\phi_s$ which satisfy the classical
equations and agree with given boundary data on $\Sigma$. The fields
can be expanded about these saddle points to obtain a perturbative
expansion of the quantum amplitude,
\begin{equation}
\Gamma=\Gamma^{(0)}+\Gamma^{(1)}+\dots
\end{equation}
where the leading term $\Gamma^{(0)}=S[\phi_s]$. The one--loop term
involves a linear operator $\Delta$ depending on the background
$\phi_s$,
\begin{equation}
\Gamma^{(1)}=\case1/2\log\det\,\Delta.
\end{equation}

Gauge fields can only be correctly accounted for if we introduce
gauge--fixing terms ${\cal F}[\phi]$ and corresponding ghost fields.
Then,
\begin{equation}
\Gamma^{(1)}=\case1/2\sum_j(-1)^{f_j}\log\det\,\Delta_j,
\end{equation}
where $j$ labels the different types of field and $f_j=1$ for fermions
and ghosts.

The determinant is infinite and has to be regulated. We shall use
$\zeta$--function regularisation with,
\begin{equation}
\zeta_j(s)=\sum_n\lambda_n^{-s},
\end{equation}
where $\lambda_n$ are the eigenvalues of the operator $\Delta_j$. This
allows us to define
\begin{equation}
\log\det\,\Delta_j=-\zeta_j'(0)-\zeta_j(0)\log\,\mu^2.
\end{equation}
where $\mu$ is the renormalisation scale.

Results on the heat kernel expansion of operators can be used to find
the scale dependent term on an arbitrary bounded manifold. (The
mathematical background can be found in \cite{gilkey}).If the heat
kernel expansion coefficients of the operator $\Delta$ on ${\cal M}$
are denoted by $B_m(\Delta,{\cal M})$, then
\begin{equation}
\zeta_j(0)=B_2(\Delta_j,{\cal M}).
\end{equation}
The renormalisation scale dependence of the path integral is therefore,
\begin{equation}
\mu{d\Gamma\over d\mu}=-\sum_j(-1)^{f_j}B_2(\Delta_j,{\cal M}).
\end{equation}
We will denote the sum by $B(\phi,{\cal M})$. In a theory that is
otherwise independent of scale this term is the conformal anomaly.

In general, the renormalisation scale dependence can be expressed in
terms of local invariants and the ceofficients are one--loop beta
functions. To be specific, consider the operator
\begin{equation}
-D^2+X
\end{equation}
where $D_a$ is a gauge derivative. The $B_2$ coefficients can be
put into the form
\begin{equation}
16\pi^2 B_2(\Delta,{\cal M})=\int_{\cal M} b_2(\Delta) d\mu +
\int_{\cal \partial M} c_2(\Delta) d\mu.
\end{equation}
Volume terms up to $b_3$ are
available in the literature. The surface terms depend upon the choice
of boundary conditions. We will use mixtures of Dirichlet and Neumann
boundary conditions,
\begin{equation}
P_-\phi=0\hbox{,~~~~~~}\left(\psi+n\cdot\nabla\right)P_+\phi=0,
\label{bc}
\end{equation}
where $P_\pm$ are projection operators and $n$ is the normal vector.

The results can be expressed in terms of polynomials in the curvature
tensor of the manifold $R_{abcd}$ and the extrinsic curvature of the
boundary $k_{ab}$,
\begin{equation}
q = {\case8/3} k^3 + {\case{16}/3} k_a^{\ b} k_b^{\ c} k_c^{\ a} - 8
kk_{ab}
k^{ab} + 4 kR - 8 R_{ab} (k n^a n^b + k^{ab}) + 8 R_{abcd} k^{ac}
n^b n^d
\end{equation}
and
\begin{equation}
g = k_a^{\ b} k_b^{\ c} k_c^{\ a} - k k_{ab} k^{ab} + {\case2/9} k^3
\end{equation}
For Dirichlet boundary conditions,
\begin{equation}
c^D_2 = {\rm tr}\left(- {\case1/{360}} q + {\case2/{35}} g - {\case1/3}
(X - {\case1/6} R) k - {\case1/2} n. \nabla (X - {\case1/6} R) +
{\case1/{15}} C_{abcd} k^{ac}n^b n^d\right),
\end{equation}
whilst for Robin boundary conditions,
\begin{eqnarray}
c_2^R =&& {\rm tr}\bigl(- {\case1/{360}}q + {\case2/{45}} g -
{\case1/3} (X - {\case1/6} R) k + {\case1/2} n.\nabla (X - {\case1/6}
R)
-{\case4/3}(\psi-{\case1/3}k)^3\\
  &&+ 2 (X - {\case1/6} R) \psi -  (\psi - {\case1/3} k)
({\case2/{45}} k^2 - {\case2/{15}} k_{ab} k^{ab}) + {\case1/{15}}
C_{abcd} k^{ac}n^b n^d\bigr).
\end{eqnarray}

For mixed boundary conditions,
\begin{eqnarray}
c_2 =&&{\rm tr} (P_+ c_2^R + P_- c_2^D - {\case2/{15}} P_{+|a}
P_+{}^{|a} \, k - {\case4/{15}} P_{+|a} P_{+|b} k^{ab}\\
&&+ {\case4/3} P_{+|a} P_+{}^{|a} P_+ \psi - {\case2/3} P_+
P_+{}^{|a} n^b F_{ab})
\end{eqnarray}
where $P_{+|a}$ denotes the surface derivative of $P_+$. Two of the
terms include corrections by Vassilevich \cite{v} of the results in
Branson and Gilkey \cite{branson}. The final term also corrects a sign
error in ref. \cite{moss2}.

Because one of the most important applications of this work is to
quantum cosmology, we will take the spacetime curvature to satisfy
vacuum Einstein equations with a cosmological constant, i.e.
$R_{ab}=\Lambda g_{ab}$. This condition removes some terms from
the results but leaves a high degree of generality. The heat kernel
coefficients will be of the form
\begin{equation}
b_2(\Delta)=\alpha_0\Lambda^2+\alpha_2R_{abcd}R^{abcd}
\end{equation}
and
\begin{equation}
c_2(\Delta)=\beta_1\Lambda k+\beta_2 k^3 +\beta_3kk_{ab}k^{ab}
+\beta_4k_a^{\ b}k_b^{\ c}k_c^{\ a}+\beta_5C_{abcd}k^{ac}n^bn^d.
\end{equation}
Special cases include the four--sphere ${\cal S}$, disc ${\cal D}$ and
spherical cap ${\cal C}$. The disc is a region of flat space bounded by
a three--sphere and the cap a region of the four--sphere with maximum
colatitude $\theta$. The heat kernel coefficients in these cases
simplify to,
\begin{equation}
B_2(\Delta,{\cal S})=\case3/2\alpha_0+4\alpha_1\hbox{,~~~~~~}
B_2(\Delta,{\cal D})=\case{27}/8\beta_2+\case9/8\beta_3+\case3/8\beta_4
\end{equation}
and
\begin{equation}
B_2(\Delta,{\cal C})=
B_2(\Delta,{\cal S})(\case1/2-\case3/4\cos\theta+\case1/4\cos^3\theta)
+B_2(\Delta,{\cal D})\cos^3\theta
+\case9/8\beta_1\cos\theta\sin^2\theta.
\end{equation}
The disc and the cap are related in the limit that $\theta=0$, then
$B_2(\Delta,{\cal C})=B_2(\Delta,{\cal D})$.

\section{DIRAC FERMIONS}

The conformal anomaly has been evaluated explicitly for Dirac fermions
on a disc \cite{death,death2,k,k2,k3}. This gives us an opportunity to
check the general results.

We will use boundary conditions that can be implemented locally by
means of projection operators. There is only one possible choice
\cite{Luckock}, $\psi=k/2$ and
\begin{equation}
P_+=\case1/2(1-\gamma_5\gamma\cdot n).
\end{equation}
(The gamma matrix conventions are such that
$\{\gamma_a,\gamma_b\}=-2g_{ab}$ and for the commutator of derivatives
$[D_a,D_b]= -\case1/8[\gamma_c,\gamma_d] R^{cd}_{\ \ ab}$).

The fermion operator is given by
\begin{equation}
\Delta=(\gamma\cdot D)^2=-D^2+\case1/4R
\end{equation}
Substitution into the general formulae gives the values shown in the
table \ref{fermions}.
In the special cases of the disc and spherical cap,
\begin{equation}
B_2(\Delta,{\cal C})=B_2(\Delta,{\cal D})=
\case1/2B_2(\Delta,{\cal S})=\case{11}/{180}
\end{equation}
and all of the $\theta$ dependence disappears.

\section{ANTISYMMETRIC TENSORS}

In four dimensions antisymmetric tensor fields $A_p$ can have rank $p$,
the number of indices, from zero to four. Fields with rank zero and one
represent scalar and vector boson fields whilst a rank two field is
locally equivalent to a scalar and ranks three and four have no
physical degrees of freedom. The classical theory of the rank
one field is conformally invariant in four dimensions but not the
others.

The Riemannian action for a p--form will be taken to be
\begin{equation}
S_p={1\over 4}\int F_{a_1\dots a_{p+1}}F^{a_1\dots a_{p+1}} d\mu
\end{equation}
where $F$ is the curvature $dA_p$. The integral is taken over the volume
of the manifold.

The action has a gauge invariance $\delta_gA_p=dA_{p-1}$, where
$A_{p-1}$ is an arbitrary antisymmetric tensor of rank $p-1$. This means
that we have to fix the gauge and introduce ghost fields. Following the
literature \cite{tm,siegel,obukhov} we use a gauge fixing condition
$\delta A_p=0$, then there are two anticommuting ghosts of rank $p-1$,
three commuting ghosts of rank $p-2$ and so on.

The anomalous rescaling of the action becomes
\begin{equation}
\mu{d \Gamma_p\over d\mu}=
-\sum_{q=0}^{p}(-1)^{p-q}(p-q+1)\log\det(\Delta_q)
\end{equation}
where $\Delta_q$ is the Hodge--de Rahm operator $d\delta+\delta d$,
which results from gauge fixing.

The boundary conditions on the ghost fields can be found by an
application of BRS symmetry. The BRS transformations are given by
\begin{equation}
\delta_{BRS} A_p=d A_{p-1}.
\end{equation}
In consequence we require a set of boundary conditions which are
preserved under exterior differentiation. A suitable set has been
described by Gilkey \cite{gilkey}. Projection operators $P_\pm$ are
defined in such a way that $P_+$ has only tangential components on the
boundary $\Sigma$, and $P_++P_-=1$. Then absolute boundary conditions
are defined by equation (\ref{bc}), with
\begin{equation}
\psi_{a_1\dots a_p}^{b_1\dots b_p}=
k_{a_1}^{\ b_1}\delta_{a_2\dots a_p}^{b_2\dots b_p}
\end{equation}

Another set of boundary conditions, called relative, can be defined by
dualising this set. Relative boundary conditions are BRS invariant when
they apply to eigenstates of the Hodge--de Rahm operator. In our
earlier work, we called the two sets of boundary conditions electric
and magnetic, depending on which fields were held fixed on the
boundary. The surface of a conductor leads to relative boundary
conditions on the electromagnetic potential.

The values of $\alpha$ and $\beta$ for $B_2(\Delta_p,{\cal M})$
obtained from the general formulae are shown in table \ref{forms}.
These results are for absolute boundary conditions. The results for
relative boundary conditions are given by the results for the dual
form.

The values of $\alpha$ and $\beta$ can be combined to produce the
conformal anomaly of the rank $p$ tensor including the ghost terms.
Results are given in table \ref{absolute} for absolute and table
\ref{relative} for relative boundary conditions.

\section{DISCUSSION}

As we mentioned in the introduction, calculations of the anomaly using
the heat kernel coefficients \cite{moss,moss2,b}
have been at odds with direct calculations. Different covariant
techniques can now be reconciled, particularly for the disc and the
spherical cap. $B_2({\cal D})$ has been calculated by a number of
people using direct calculations
\cite{death,death2,k2} for majorana fermions with the
local boundary conditions described above. They find that
$B_2({\cal D})=\case{11}/{180}$ which agrees with the result reported
here. For fermions $B_2({\cal C})$ has also been calculated
directly by \cite{k2,k3} who find that $B_2({\cal C})=\case{11}/{180}$,
which is again in agreement with our calculations.
Finally, in a recent paper \cite{e2} Esposito and Kamenshchik have
calculated the anomaly on a disc for the one--form including the ghosts.
They find the result $-\case{31}/{90}$ (as we do). This still disagrees
with a calculation using only the physical degrees of freedom, but they
argue that the difference arises from the difficulty in defining a
canonical decomposition for the disc.

An examination of tables reveals some interesting coincidences.
The boundary terms for ranks three and four combine with the
volume terms by the Gauss--Bonnet theorem and form an integral expression
for the Euler number $\chi$ on an arbitrary manifold. Let us denote the
combined heat kernel coefficients by $B^a(A_p,{\cal M})$ and
$B^r(A_p,{\cal M})$, for absolute and relative
boundary conditions, then
\begin{eqnarray}
&B^a(A_2,{\cal M})=B^r(A_0,{\cal M})+\chi\hbox{,~~~~~}
&B^r(A_2,{\cal M})=B^a(A_0,{\cal M})+\chi\\
&B^a(A_3,{\cal M})=B^r(A_3,{\cal M})=-2\chi,\
&B^a(A_4,{\cal M})=B^r(A_4,{\cal M})=3\chi.
\end{eqnarray}

These relationships can be derived directly using standard cohomology
theory and a new identity
\begin{equation}
2B_2(\Delta_4,{\cal M})-B_2(\Delta_3,{\cal M})
+B_2(\Delta_2,{\cal M})-2B_2(\Delta_0,{\cal M})=0,
\end{equation}
for absolute and relative boundary conditions separately. This same
identity is also responsible for the fact that the vector gauge field
results are independent of the choice of boundary conditions. The last
identity is unlikely to be coincidental and indicates a deeper
underlying structure.

\begin{table}
\begin{tabular}{crrrrrrrr}
$\alpha_0$&$\alpha_1$&$\beta_1$&$\beta_2$&
$\beta_3$&$\beta_4$&$\beta_5$&$B_2({\cal S})$&$B_2({\cal D})$\\
\hline
$\case2/{15}$&$-\case7/{360}$&$\case{11}/{135}$&$\case{17}/{945}$&
$\case{13}/{315}$&$-\case{116}/{945}$&$-\case7/{45}$&$\case{11}/{90}$&
$\case{11}/{180}$\\
\end{tabular}
\caption{Coefficients for the conformal anomaly of a Dirac fermion}
\label{fermions}
\end{table}

\begin{table}
\begin{tabular}{crrrrrrr}
rank&$\alpha_0$&$\alpha_1$&$\beta_1$&$\beta_2$&
$\beta_3$&$\beta_4$&$\beta_5$\\
\hline
$0$&$\case1/5$&$\case1/{180}$&$\case{29}/{135}$&$\case1/{27}$&
$\case1/{45}$&$\case4/{135}$&$\case2/{45}$\\
\hline
$1$&$\case2/{15}$&$-\case{11}/{180}$&$-\case4/{135}$&
$-\case{268}/{945}$&$\case{304}/{315}$&$-\case{76}/{189}$&
$-\case{22}/{45}$\\
\hline
$2$&$-\case2/{15}$&$\case{33}/{90}$&$\case{38}/{45}$&
$\case{208}/{315}$&$-\case{76}/{35}$&$\case{572}/{315}$&
$\case{132}/{45}$\\
\hline
$3$&$\case2/{15}$&$-\case{11}/{180}$&$-\case4/{135}$&
$-\case{328}/{945}$&$\case{268}/{315}$&$-\case{356}/{945}$&
$-\case{22}/{45}$\\
\hline
$4$&$\case1/{5}$&$\case{1}/{180}$&$\case{29}/{135}$&
$\case{1}/{189}$&$-\case{11}/{315}$&$\case{8}/{189}$&
$\case{2}/{45}$\\
\end{tabular}
\caption{Coefficients for terms in the $B_2$ coefficient for p-forms
with
absolute boundary conditions}
\label{forms}
\end{table}

\begin{table}
\begin{tabular}{crrrrrrrrr}
 rank&$\alpha_0$&$\alpha_1$&$\beta_1$&$\beta_2$&
$\beta_3$&$\beta_4$&$\beta_5$&$B({\cal S})$&$B({\cal D})$\\
\hline
$0$&$\case1/{5}$&$\case1/{180}$&$\case{29}/{135}$&$\case1/{27}$&
$\case1/{45}$&$\case4/{135}$&$\case2/{45}$&$\case{29}/{90}$
&$\case{29}/{180}$\\
\hline
$1$&$-\case4/{15}$&$-\case{13}/{180}$&$-\case{62}/{135}$&
$-\case{338}/{945}$&$\case{58}/{63}$&$-\case{436}/{945}$&
$-\case{26}/{45}$&$-\case{31}/{45}$&$-\case{31}/{90}$\\
\hline
$2$&$\case1/{5}$&$\case{91}/{180}$&$\case{209}/{135}$&
$\case{253}/{189}$&$-\case{1271}/{315}$&$\case{512}/{189}$&
$\case{182}/{45}$&$\case{209}/{90}$&$\case{179}/{180}$\\
\hline
$3$&$0$&$-1$&$-\case8/3$&
$-\case8/3$&$8$&$-\case{16}/3$&$-8$&$-4$&$-2$\\
\hline
$4$&$0$&$\case3/2$&$4$&$4$&$-12$&$8$&$12$&$6$&$3$\\
\end{tabular}
\caption{Coefficients for terms in the conformal anomaly or scaling
terms for
p-forms with absolute boundary conditions including ghosts. $B_2({\cal
S})$
refers to the combination of $B_2$ coefficients on a sphere and
$B_2({\cal D})$ on a disk.}
\label{absolute}
\end{table}

\begin{table}
\begin{tabular}{crrrrrrrrr}
rank&$\alpha_0$&$\alpha_1$&$\beta_1$&$\beta_2$&
$\beta_3$&$\beta_4$&$\beta_5$&$B({\cal S})$&$B({\cal D})$\\
\hline
$0$&$\case1/{5}$&$\case1/{180}$&$\case{29}/{135}$&
$\case1/{189}$&$-\case{11}/{315}$&$
\case8/{189}$&$\case2/{45}$&$\case{29}/{90}$&$-\case1/{180}$\\
\hline
$1$&$-\case4/{15}$&$-\case{13}/{180}$&$-\case{62}/{135}$&
$-\case{338}/{945}$&$\case{58}/{63}$&$-\case{436}/{945}$&
$-\case{26}/{45}$&$-\case{31}/{45}$&$-\case{31}/{90}$\\
\hline
$2$&$\case1/{5}$&$\case{91}/{180}$&$\case{209}/{135}$&
$\case{37}/{27}$&$-\case{1253}/{315}$&$\case{364}/{135}$&
$\case{182}/{45}$&$\case{209}/{90}$&$\case{209}/{180}$\\
\hline
$3$&$0$&$-1$&$-\case8/3$&$-\case8/3$&$
8$&$-\case{16}/{3}$&$-8$&$-4$&$-2$\\
\hline
$4$&$0$&$\case{3}/{2}$&$4$&$4$&$-12$&$8$&$12$&$6$&$3$\\
\end{tabular}
\caption{Coefficients for terms in the conformal anomaly or scaling
terms for
p-forms with relative boundary conditions including ghosts. $B_2({\cal
S})$
refers to the combination of $B_2$ coefficients on a sphere and
$B_2({\cal D})$ on a disk.}
\label{relative}
\end{table}

\end{document}